# Beam-commissioning-oriented optics study of HFRS Phase-I based on measured magnetic field data

Ke Wang [a, b, c, d], Li-Na Sheng [b, c*], Xue-Heng Zhang [b, c], Bei-Min Wu [b, c], Ming-Bang Lü[b, c], Dong-Sheng Ni [b, c], Jing Yang [b, c], Xiang Zhang [b, c], Fu-Qiang Liu [b, c], Qing-Gao Yao [b, c], Xiao-Wei Xu [b, c], Ya-Jun Zheng [b, c], Guo-Dong Shen [b, c], Geng Wang [b, c], You-Jin Yuan [b, c], Jian-Cheng Yang [b, c], Liang Lu [a, d†]

a. Sino-French Institute of Nuclear Engineering and Technology, Sun Yat-sen University, Zhuhai 519082, P. R. China

b. Institute of Modern Physics, Chinese Academy of Sciences, Lanzhou 730000, P. R. China

c. University of Chinese Academy of Sciences, Beijing 100049, P. R. China

d. United Laboratory of Frontier Radiotherapy Technology of Sun Yat-sen University & Chinese Academy of Sciences Ion Medical Technology Co., Ltd, Guangzhou 510030, P. R. China

**Abstract**

The construction of the first phase of the High energy FRagment Separator (HFRS Phase-I) has already been completed and it is anticipated to start beam commissioning in autumn 2025. This paper presents the first order and higher order beam optics calculations for the HFRS Phase-I, based on measured magnet data, and evaluates its experimental performance in preparation for beam commissioning. The first order optics of HFRS is calculated based on the sliced magnetic fields and the higher order aberrations are corrected using a self-compiled program. Monte Carlo particle tracking is employed to analyze the beam phase spaces on the focal planes, and a thorough examination of these phase spaces demonstrates that the higher order aberrations have been well corrected. Moreover, the experimental performance of HFRS is evaluated based on the corrected higher order optics, yielding satisfactory results: the secondary beams of interest can be well separated and exhibit high transmission efficiency. This work provides valuable insights for the upcoming beam commissioning of HFRS Phase-I. The effective correction of higher order aberrations and optimized magnet settings lay a solid foundation for future experiments.

**Keywords**: radioactive beam, in-flight separator, beam optics, aberration correction, HFRS

[*] Corresponding author: Li-Na Sheng, Email: shenglina@impcas.ac.cn.

[†] Corresponding author: Liang Lu, Email: luliang3@mail.sysu.edu.cn.

## 1. Introduction

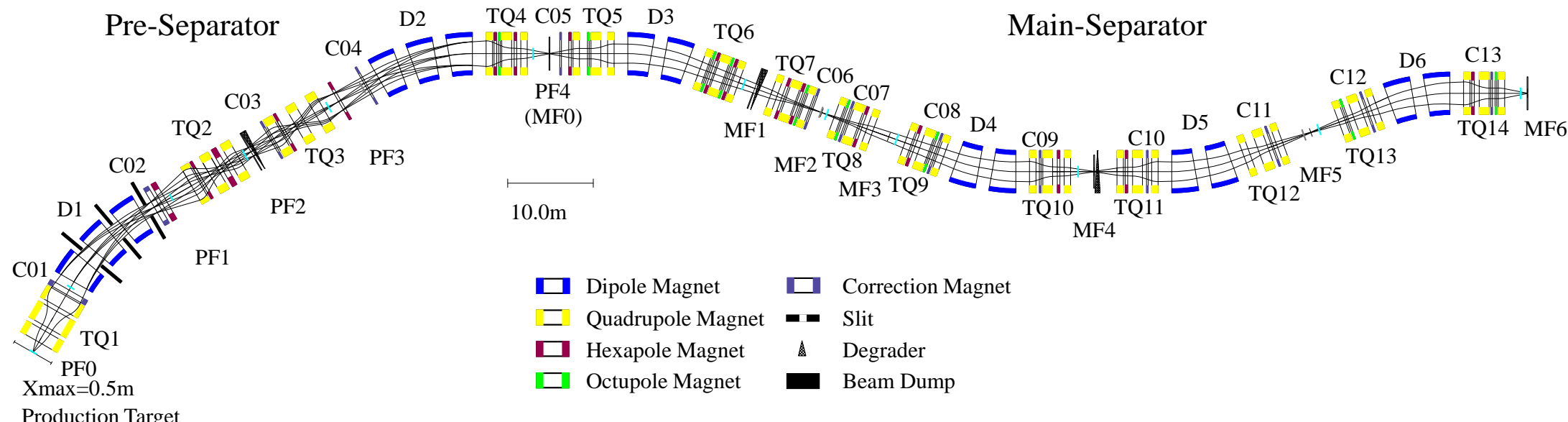

Figure 1. Layout of HFRS.

**H**igh energy **FR**agment **S**eparator (HFRS) [1, 2], as an important experiment terminal of the **H**igh **I**ntensity heavy-ion **A**ccelerator **F**acility (HIAF) [3-5], is a two-stage radioactive beam in-flight separator consisting of pre-separator and main-separator, as shown in Fig. 1. The total length of HFRS is 191.38 m. It uses the $B\rho$-$\Delta E$-$B\rho$ method [6] to spatially separate the radioactive isotopes produced at PF0 by projectile fragmentation or fission of heavy ions. The booster ring (BRing) [7, 8] of HIAF is capable of accelerating a wide range of ions, from light ones like protons to heavy ones such as $^{238}$U. Table 1 lists the main parameters of several primary beams. HFRS can operate in two modes: the dispersive mode and the achromatic mode. The construction of HFRS is carried out in two phases. The maximum magnetic rigidity ($B\rho$) of HFRS Phase-I is 13.2 Tm, and it will be enhanced to 25 Tm in Phase-II. Compared to the earlier design [1], the main change in the Phase-I is to replace all superconducting discrete cosine theta multipole magnets [9] with normal-conducting multipole magnets. Table 2 lists the main parameters of Phase-I and Phase-II of HFRS.

Table 1. Main parameters of several primary beams provided by BRing.

| Ion | Intensity (ppp) | Energy (MeV/u) |
| --- | --- | --- |
| $^{238}U^{34+}$ | 1.00E+11 | 800 |
| $^{129}Xe^{27+}$ | 1.80E+11 | 1400 |
| $^{78}Kr^{19+}$ | 3.00E+11 | 1750 |
| $^{40}Ar^{12+}$ | 5.00E+11 | 2300 |
| $^{18}O^{6+}$ | 6.00E+11 | 2600 |
| p | 2.00E+12 | 9300 |

ppp: particles per pulse.

Table 2. Main parameters of HFRS in Phase-I and Phase-II.

| Parameters | Phase I | Phase II |
| --- | --- | --- |
| $B\rho_{max}$ [Tm] | 13.2 | 25 |
| Beam size at target [mm] | ±1(H) / ±1.5(V) | |
| Angular acceptance [mrad] | ±30(H) / ±25(V) | |
| Momentum acceptance | ±2% | |
| Ion-optical mode | Dispersive mode, Achromatic mode | |
| First order Momentum resolution | Dispersive mode: 800(Pre) / 700(Main)<br>Achromatic mode: 800(Pre) / 1200(Main) | |

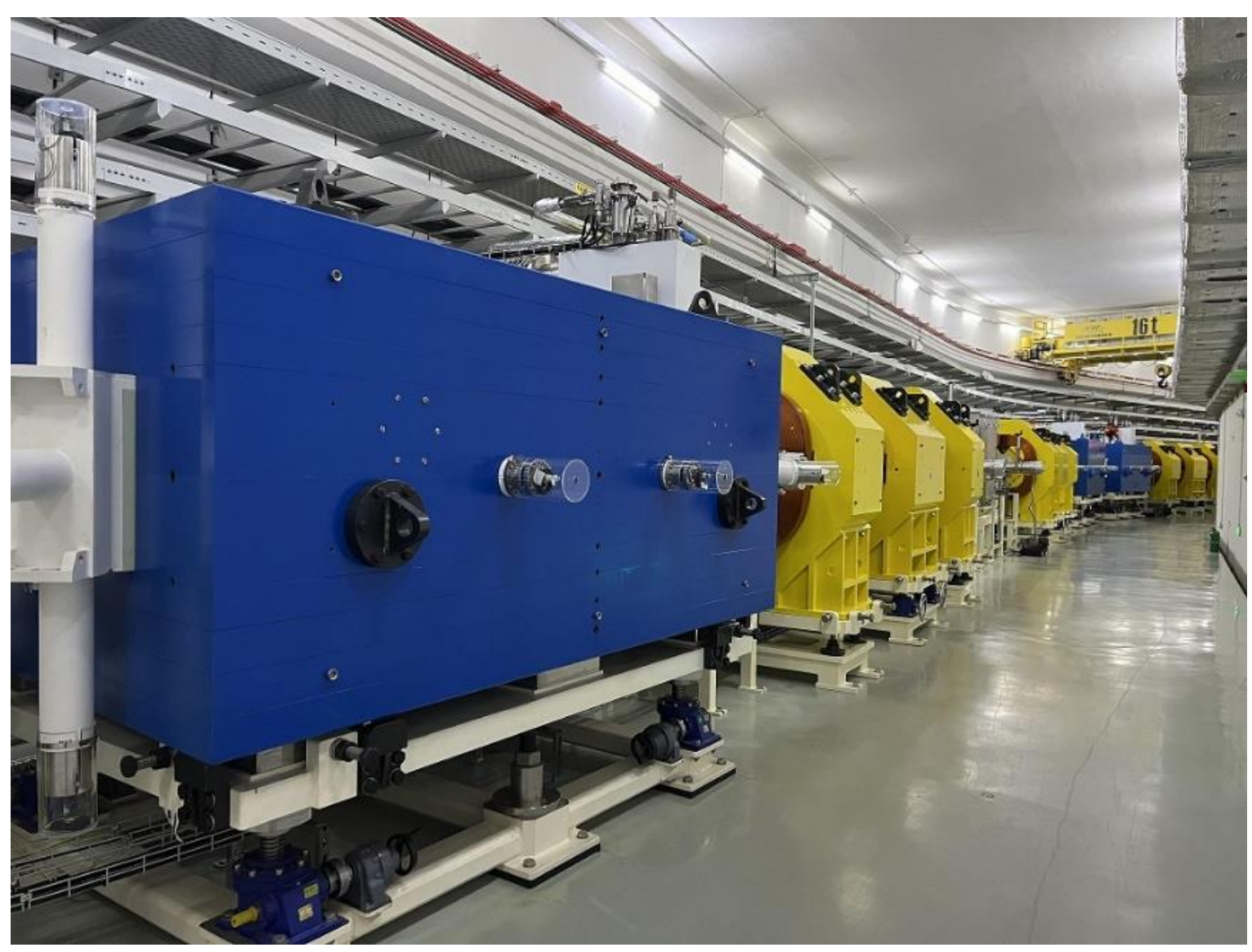

Figure 2. A partial view of the HFRS tunnel.

All the equipment of HFRS Phase-I has been installed, and the joint offline-commissioning of magnets, power supplies, beam diagnostics and vacuum has been completed, which can meet the design parameters. The beam commissioning will start in autumn 2025, and then a day-one experiment will be performed. A partial view of the HFRS tunnel is shown in Fig. 2.

In order to prepare for the upcoming beam commissioning and day-one experiment, the accurate calculation of beam optics and evaluation of experimental performance of HFRS Phase-I are presented in this paper. The beam optics calculation, including the first order sliced optics and the high order aberration correction are all based on measured magnetic field data. Therefore, the fringe fields, effective length, magnetic field homogeneity, multipole field components and so on for real magnets are all taken into account. Ion-optics calculations are performed with MAD-X [10] and COSY INFINITY [11-13] (COSY), and the experimental performance evaluation is carried out by the Monte Carlo program MOCADI [14, 15]. Then, the corresponding excitation currents of quadrupole magnets, hexapole and octupole magnets are calculated according to their I-GL curves and these simulation results. The I-GL curve is the integral of the magnetic field gradient along the reference orbit as a function of the excitation current. At present, optics calculation programs and the measured magnetic field data have been integrated into the physics control system.

## 2. Measured Fields

The magnet system of HFRS is composed of 95 magnets of various types, which includes 14 dipole magnets and 42 quadrupole magnets. There are two types of dipole magnets, including 3 normal-conducting radiation-resistant dipole magnets and 11 superferric dipole magnets. Quadrupole magnets are also two types, 3 normal-conducting radiation-resistant magnets and 39 normal-conducting magnets. Besides, 13 pairs of steerers, 17 hexapole magnets and 9 octupole magnets are adopted separately for beam position correction and higher order aberration correction. The magnet parameters of HFRS Phase-I are listed in Table 3.

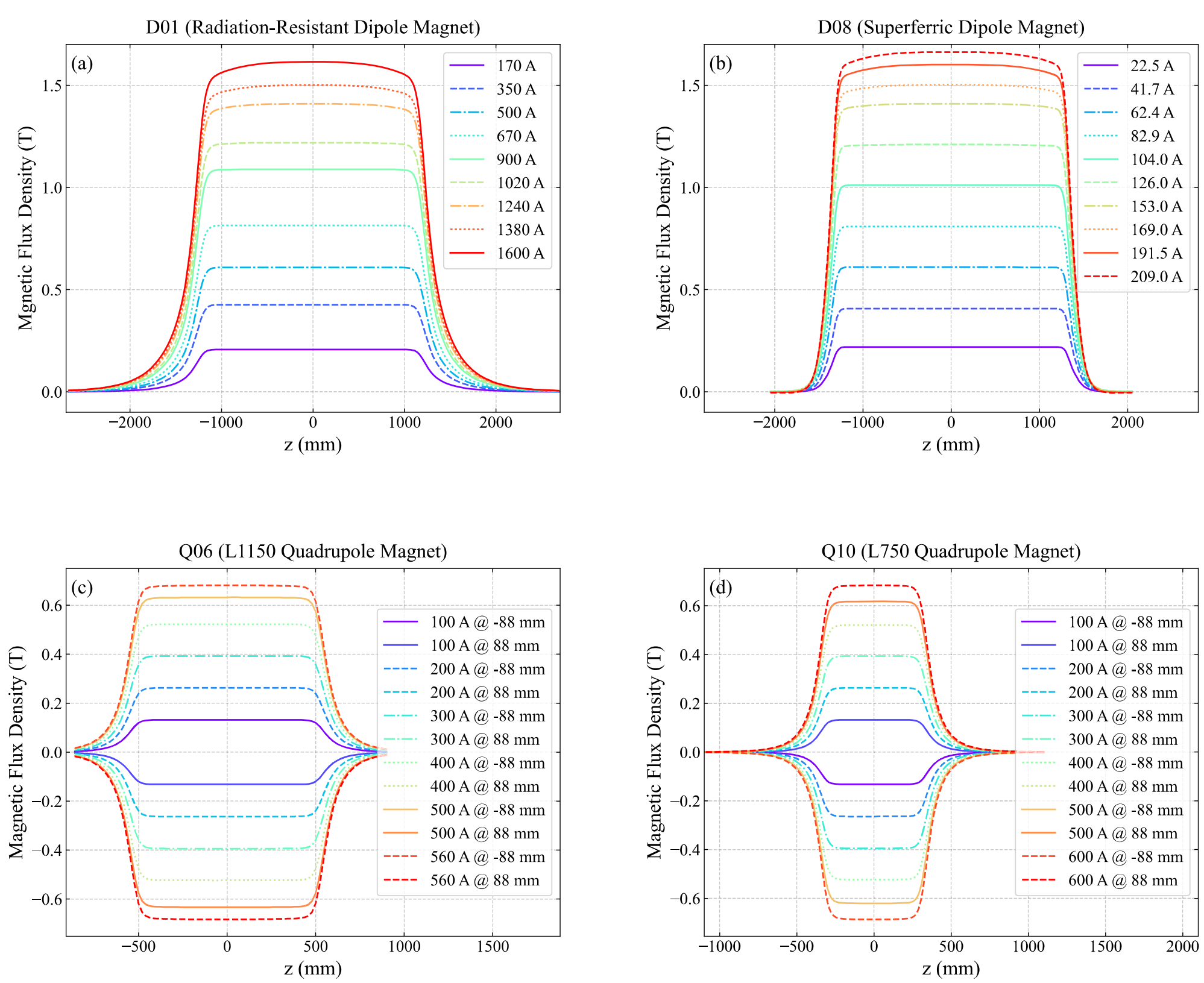

Figure 3. Longitudinal magnetic fields distribution with different excitation currents.

At present, all magnets have been tested and installed in tunnel. Figure 3 shows four typical longitudinal magnetic field distributions at the midplane through point measurement with different excitation currents, which serve as the basis for extracting fringe field information. Changes in the effective lengths and excitation curves of dipole magnets are presented in Fig. 4. The transverse fields for two types of dipole magnets are also given in Fig. 5, which can be used to describe the midplane radial fields [16]. The midplane radial field refers to the magnetic field distributed radially on the midplane of the magnet's gap, with the origin in Fig. 5 located at the center of the magnet's gap. The coordinate system used in this paper is a right-handed system, where the positive direction of the z-axis points to the direction of the beam flight, the positive direction of the x-axis points to the left of the beam, and the positive direction of the y-axis points upward.

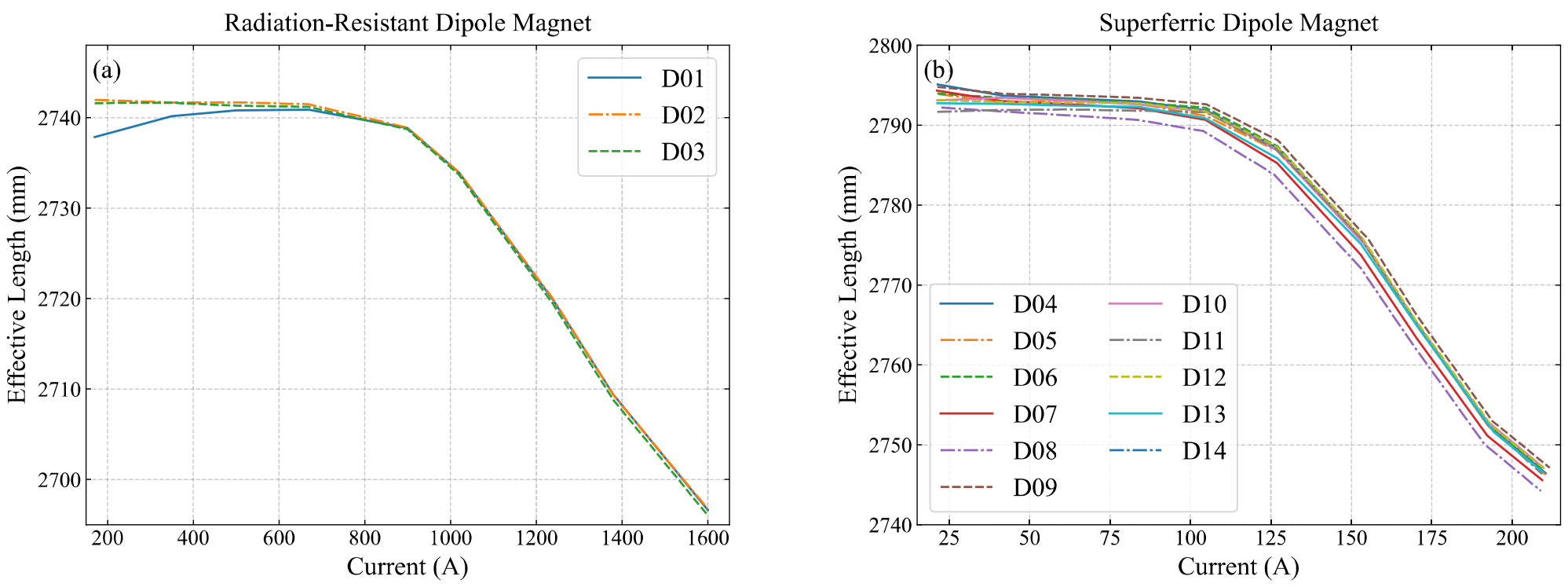

Figure 4. Effective lengths of dipole magnets.

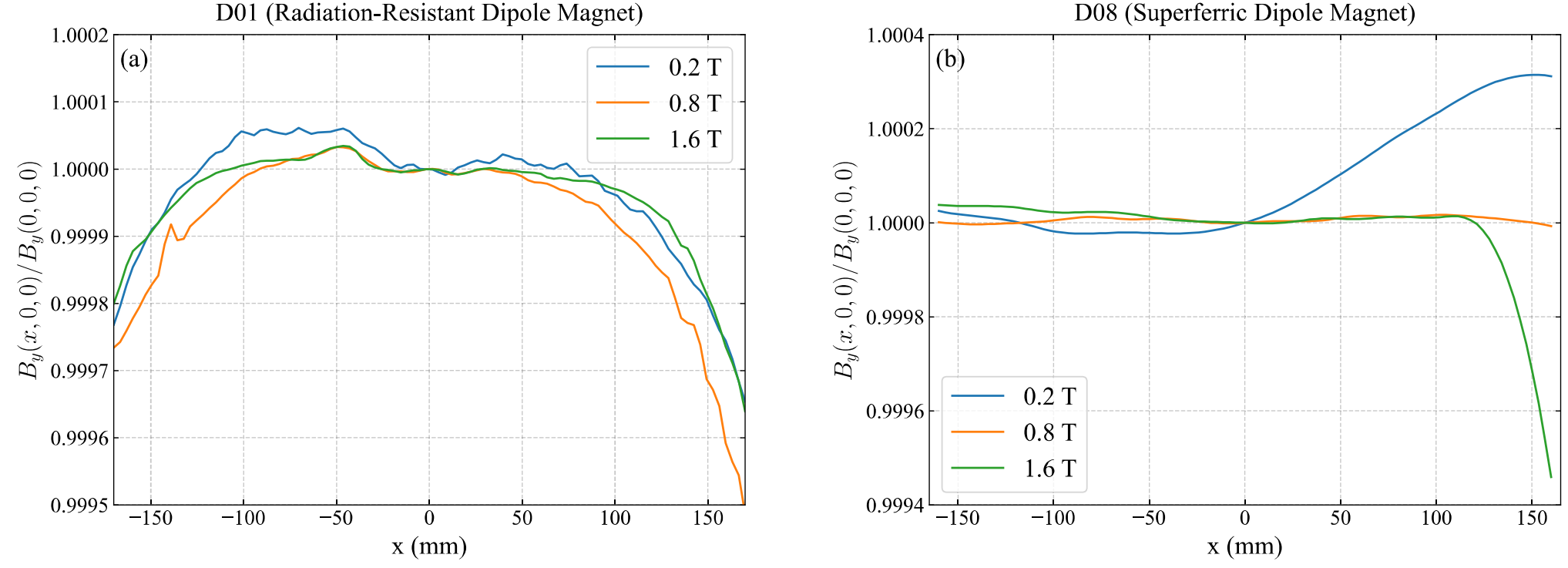

Figure 5. Transverse magnetic fields distribution of two types dipole magnets.

Table 3. Parameters of magnets of HFRS Phase-I.

| Type | Length [m] | Aperture [mm] | Max. field/ pole-tip field [T] | Bending Angle |
|---|---|---|---|---|
| Radiation-resistant dipole magnet | 2.74 | ±160 (H), ±90 (V) | 1.60 | 10° |
| Superferric dipole magnet | 2.74 | ±160 (H), ±62 (V) | 1.60 | 10° |
| Radiation-resistant quadrupole magnet type #1 | 1.50 | 200 | 0.96 | |
| Radiation-resistant quadrupole magnet type #2 | 2.20 | 274 | 1.19 | |
| Radiation-resistant quadrupole magnet type #3 | 1.60 | 286 | 0.92 | |
| Normal-conducting quadrupole magnet type #1 | 1.00 | 250 | 0.83 | |
| Normal-conducting quadrupole magnet type #2 | 1.15 | 220 | 0.85 | |
| Normal-conducting quadrupole magnet type #3 | 0.75 | 220 | 0.88 | |
| Radiation-resistant sextupole magnet type #1 | 0.60 | 270 | 0.50 | |
| Radiation-resistant sextupole magnet type #2 | 0.60 | 270 | 0.36 | |
| Sextupole magnet | 0.30 | 310 | 0.50 | |
| Octupole magnet | 0.20 | 310 | 0.35 | |
| Radiation-resistant correction magnet | 0.60 | ±232 (H), ±148 (V) | 0.04 | |
| Correction magnet | 0.20 | ±185 (H), ±90 (V) | 0.08 | |

The Enge function [17] is used for fringe field calculation, as listed in equation (1). In equation (1), $F(z)$ represents the normalized magnetic field, and $D$ represents the aperture of quadrupole magnet or gap of dipole magnet. Figure 6 shows the Enge coefficients extracted from the longitudinal field distributions of radiation-resistant dipole magnets and the superferric dipole magnet. Notably, the Enge coefficients exhibit consistent trends for different magnet types.

$$F(z)=\frac{1}{1+\exp(a_1+a_2\cdot(z/D)+\cdots+a_6\cdot(z/D)^5)}. \quad (1)$$

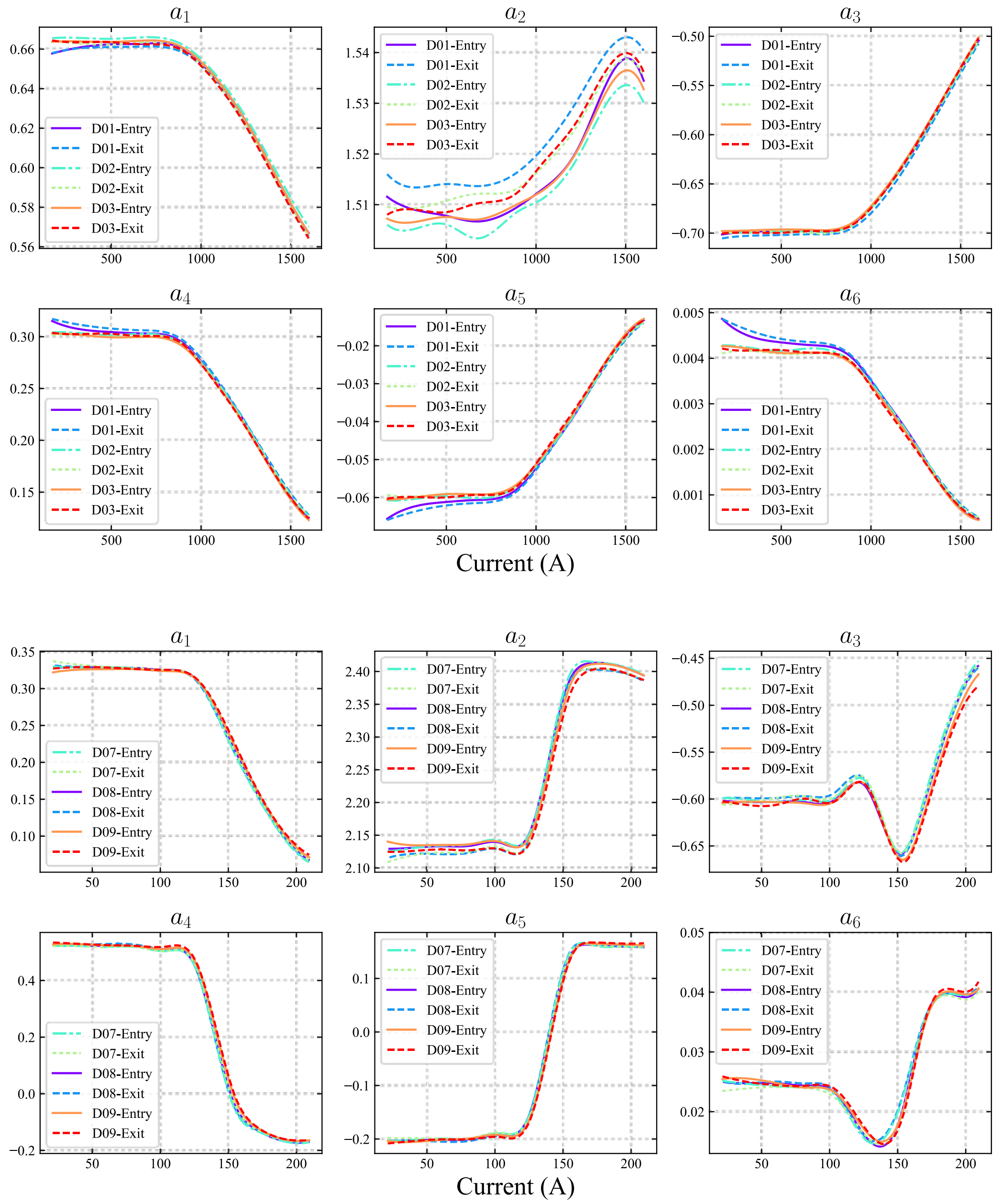

Figure 6. Enge coefficients of the fringe fields of the radiation-resistant dipole magnets (D01/02/03, upper) and the superferric dipole magnets (D07/08/09, lower) as functions of excitation currents.

The multipole field components of magnets in HFRS are obtained using the harmonic coil measurement method [18]. Figure 7 shows the multipole field components of quadrupole magnets Q06 and Q10, which are both less than 6.0E-4.

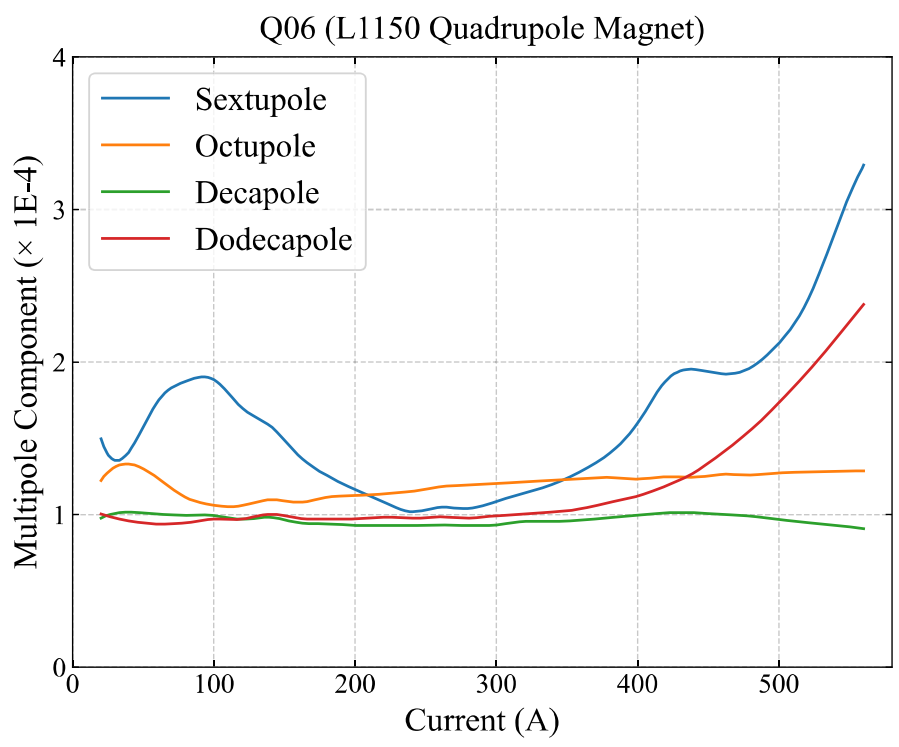

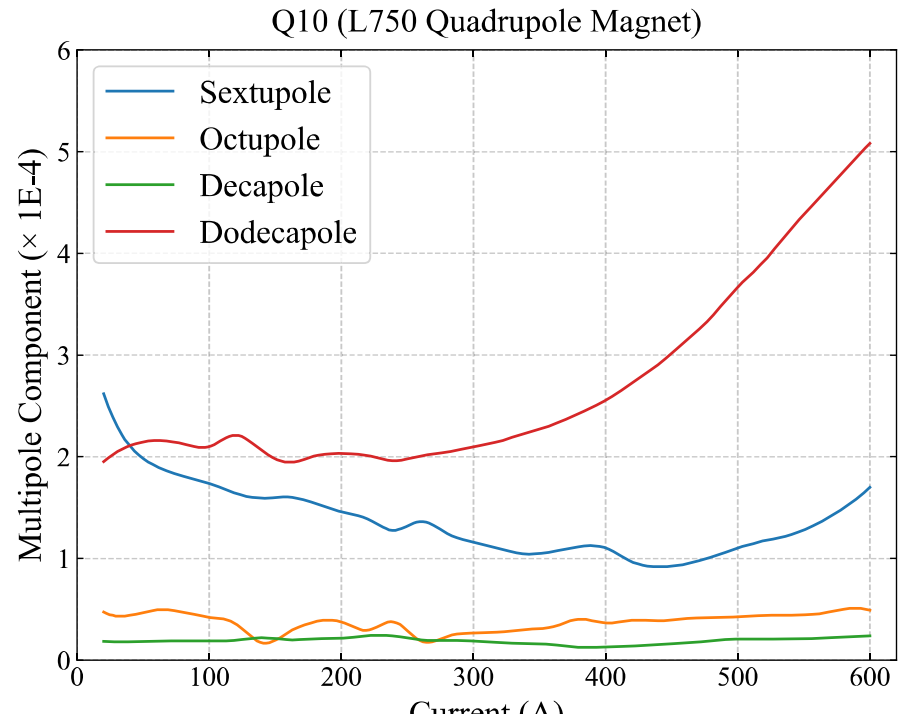

Figure 7. Multipole field components of two types of quadrupole magnets as functions of excitation currents.

## 3. Beam optics calculation

### 3.1. First order optics with sliced fields

In order to calculate the beam optics with high precision, each magnet is divided into several segments based on the measured magnetic field, which is named sliced beam optics. The detailed procedure is described in reference [7]. In this way, both the superposition fields between adjacent magnets and the fringe fields of the magnets can be taken into consideration. The first order beam optics is implemented with MAD-X and COSY INFINITY. Figure 8 shows the comparison of betatron functions and horizontal dispersion functions between the cases using measured sliced magnetic field and the hard-edge field, in dispersive mode and achromatic mode in the first order optics calculations. By adjusting the focusing strength of the quadrupole magnets, the sliced beam optics with measured field is consistent with that under the hard-edge magnetic field, and the Twiss parameters on each focal plane can reach the design goals. Then, the calculated normalized gradients of the quadrupole magnets will serve as the basis for beam commissioning. This work is very beneficial for beam commissioning, because each magnet is divided into small slices of approximately 10 mm, and these slices are still treated as a whole in the calculation.

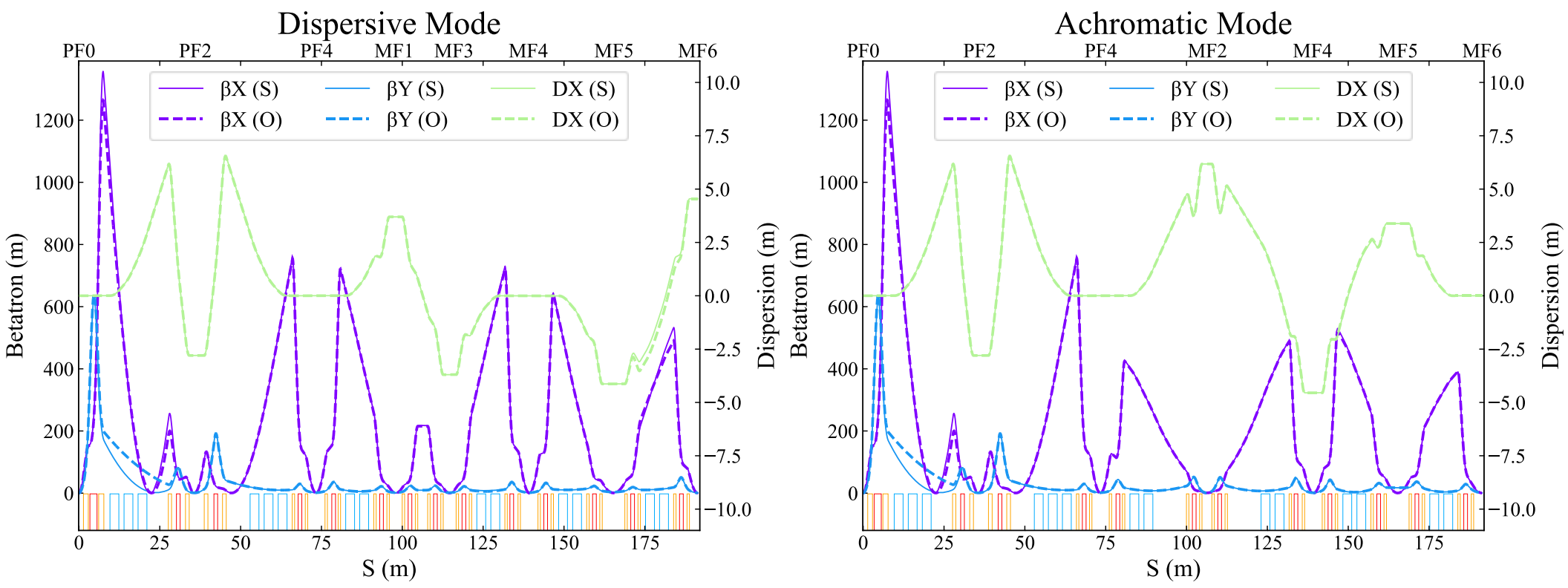

Figure 8. Comparison of betatron functions and horizontal dispersion functions for measured sliced

magnetic field and hard-edge magnetic field in two ion-optics modes. “S” represents the sliced optics. “O” represents the original optics

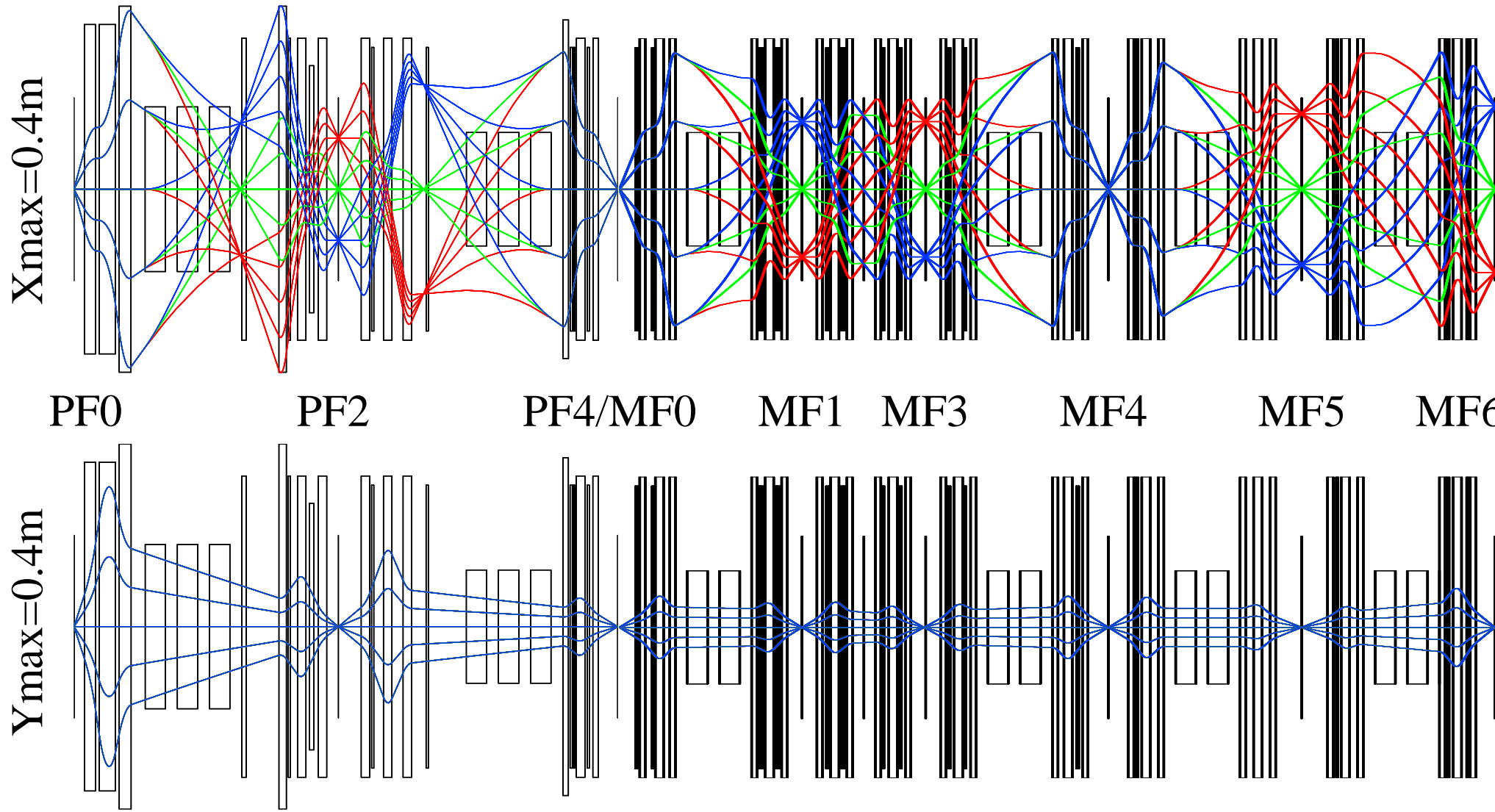

Figure 9. First order ion trajectories of HFRS in dispersive mode in the horizontal (upper) and vertical (lower) directions. The ions trajectories are drawn with $x_0$= ±1 mm, $y_0$=±1.5 mm, $a_0$=±30, ±15 and 0 mrad, $b_0$=±25, ±12.5 and 0 mrad, and Δp/p=0 and ±2%.

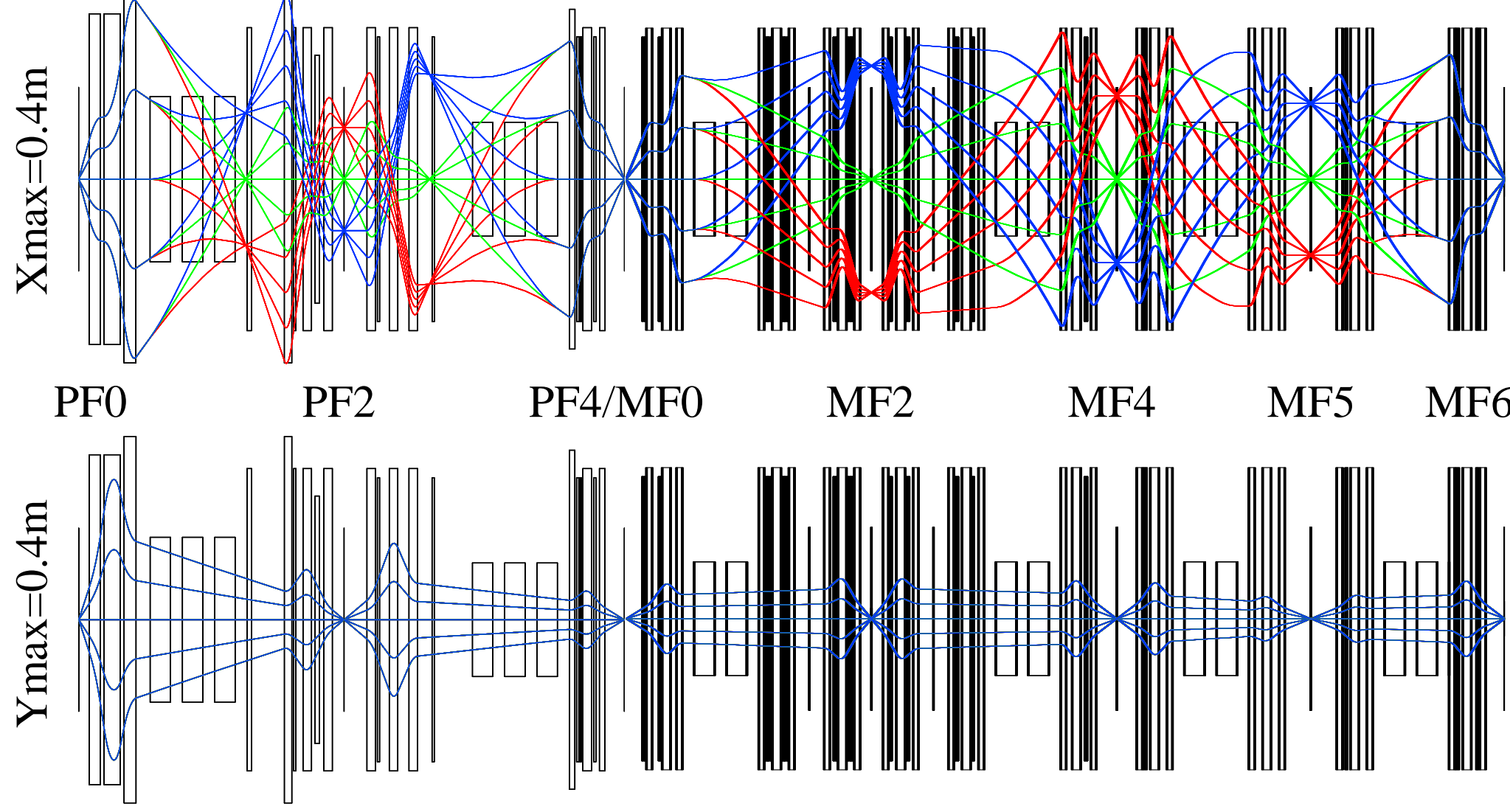

Figure 10. First order ion trajectories of HFRS in archromatic mode in the horizontal (upper) and vertical (lower) directions. The ions trajectories are drawn with $x_0$= ±1 mm, $y_0$=±1.5 mm, $a_0$=±30, ±15 and 0 mrad, $b_0$=±25, ±12.5 and 0 mrad, and Δp/p=0 and ±2%.

Table 4. Transfer matrix elements at main focal planes of HFRS in the first order ion optics, calculated by COSY INFINITY.

| | (x, x) | (x, a) | (x, δ) | (a, δ) | (y, y) | (y, b) |
|---|---|---|---|---|---|---|
| PF2 | 1.62 | 0 | -2.80 | 0 | -3.65 | 0 |
| PF4 | 1.50 | 0 | 0 | 0 | 5.00 | 0 |
| Dispersive mode | | | | | | |
| MF1 | -2.63 | 0 | 3.70 | 0 | -5.00 | 0 |
| MF3 | 2.63 | 0 | -3.70 | 0 | 5.00 | 0 |
| MF4 | -1.50 | 0 | 0 | 0 | -5.00 | 0 |
| MF5 | 3.30 | 0 | -4.13 | 0 | 7.67 | 0 |
| MF6 | -1.84 | 0 | 4.55 | 0 | -3.76 | 0 |
| Achromatic mode | | | | | | |
| MF2 | -5.30 | 0 | 6.18 | 0 | -2.57 | 0 |
| MF4 | 1.89 | 0 | -4.55 | 0 | 4.08 | 0 |
| MF5 | -3.39 | 0 | 4.13 | 0 | -7.01 | 0 |
| MF6 | 1.50 | 0 | 0 | 0 | 4.08 | 0 |

The first order beam trajectories of HFRS calculated by COSY are shown in Figs. 9 and 10. The ions trajectories are drawn with $x_0$= ±1 mm, $y_0$=±1.5 mm, $a_0$=±30, ±15 and 0 mrad, $b_0$=±25, ±12.5 and 0 mrad, and Δp/p=0 and ±2%. The transfer matrix elements of HFRS are listed in Table 4. The momentum resolution of the pre-separator at PF2 is about 800. For the main-separator, the momentum resolution is about 700 at MF1 in dispersive mode and 1200 at MF4 in achromatic mode. In Table 4, the units of the lengths and angles are meter and radian, respectively. $a=p_x/p_0$ and $b=p_y/p_0$, which are the ratios of the x-component and y-component of the total momentum $p_0$ to $p_0$. a and b are approximately equal to the divergence angles in the x and y directions (i.e., x' and y'). δ denotes momentum dispersion.

3.2. Higher order optics with aberration correction

A self-compiled program based on COSY/GICOSY code [13, 19], MOCADI code and Optuna [20] is developed for third order aberration correction. COSY and GICOSY runtime libraries are built to call COSY or GICOSY code as needed. And the calculations can be chosen whether involving the MOCADI code. The program supports multiprocessing computation and uses SQLite [21] to store computation data, which can support checkpoint to restart. The optics calculations also adopt the Enge coefficients for the fringe fields and the multipole field components of magnets, as well as the radial field coefficients.

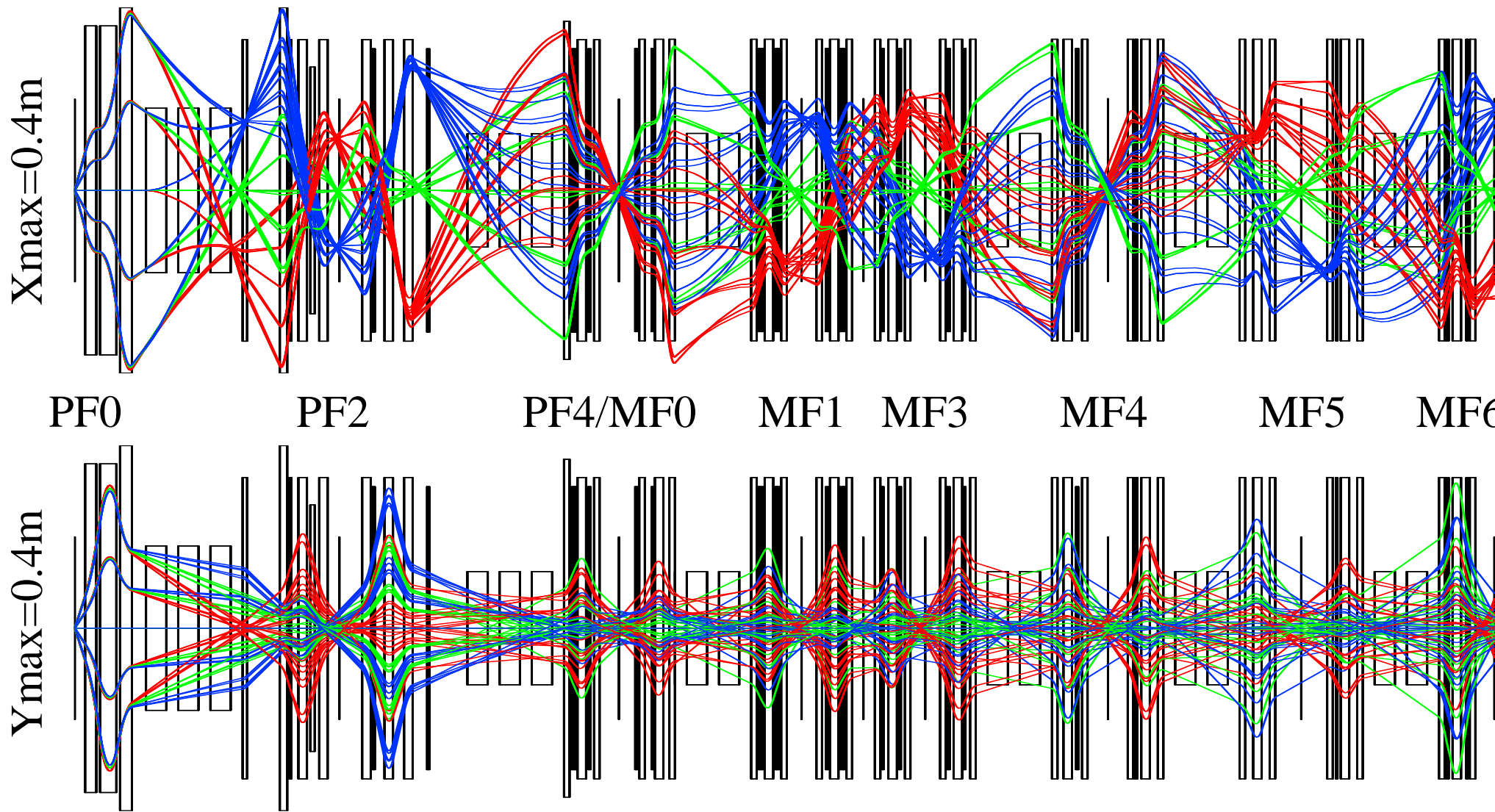

Figure 11. Ion trajectories of HFRS in dispersive mode in the horizontal (upper) and vertical (lower) directions calculated up to 5th-order. Ion trajectories are shown with $x_0$=±1 mm, $y_0$=±1.5 mm, $a_0$=±30, ±15 and 0 mrad, $b_0$=±25, ±12.5 and 0 mrad, and $\Delta p/p$=±2%.

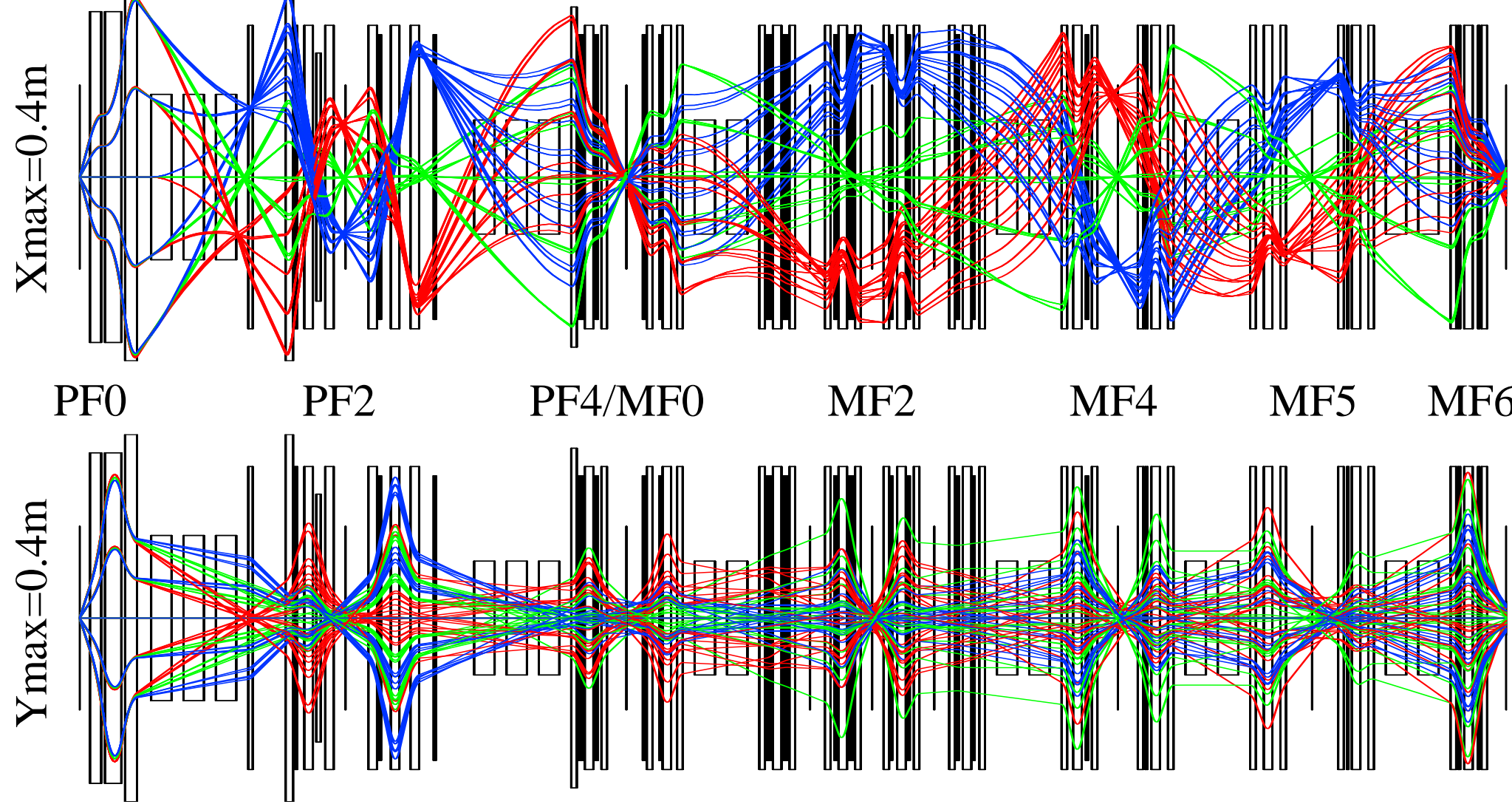

Figure 12. Ion trajectories of HFRS in achromatic mode in the horizontal (upper) and vertical (lower) directions calculated up to 5th-order. Ion trajectories are shown with $x_0$=±1 mm, $y_0$=±1.5 mm, $a_0$=±30, ±15 and 0 mrad, $b_0$=±25, ±12.5 and 0 mrad, and $\Delta p/p$=±2%.

The beam trajectories of HFRS calculated up to 5th-order without and with aberration correction are shown in Figs. 11 and 12. The ions trajectories are drawn with $x_0$= ±1 mm, $y_0$=±1.5 mm, $a_0$=±30, ±15 and 0 mrad, $b_0$=±25, ±12.5 and 0 mrad, and $\Delta p/p$=0 and ±2%. The beam phase spaces at focal planes are simulated with 300,000 particles by Monte Carlo ion tracking with COSY. The beam parameters are the same as those shown in Table 2 with a Gaussian distribution. Figures 13 and 14 show the beam phase space at PF2 and PF4 focal planes without and with third order aberration correction, respectively. Figures 15 and 16 show the beam phase space at crucial focal planes of main-separator on dispersive mode and achromatic mode without and with the third order aberration correction, respectively. As it is seen, the aberrations at these focal planes have been greatly improved after correction. Smaller aberrations are beneficial for maintaining transmission efficiency and momentum resolution power. The magnetic field strengths of the hexapole and octupole magnets obtained through higher order aberration corrections will also serve the beam commissioning.

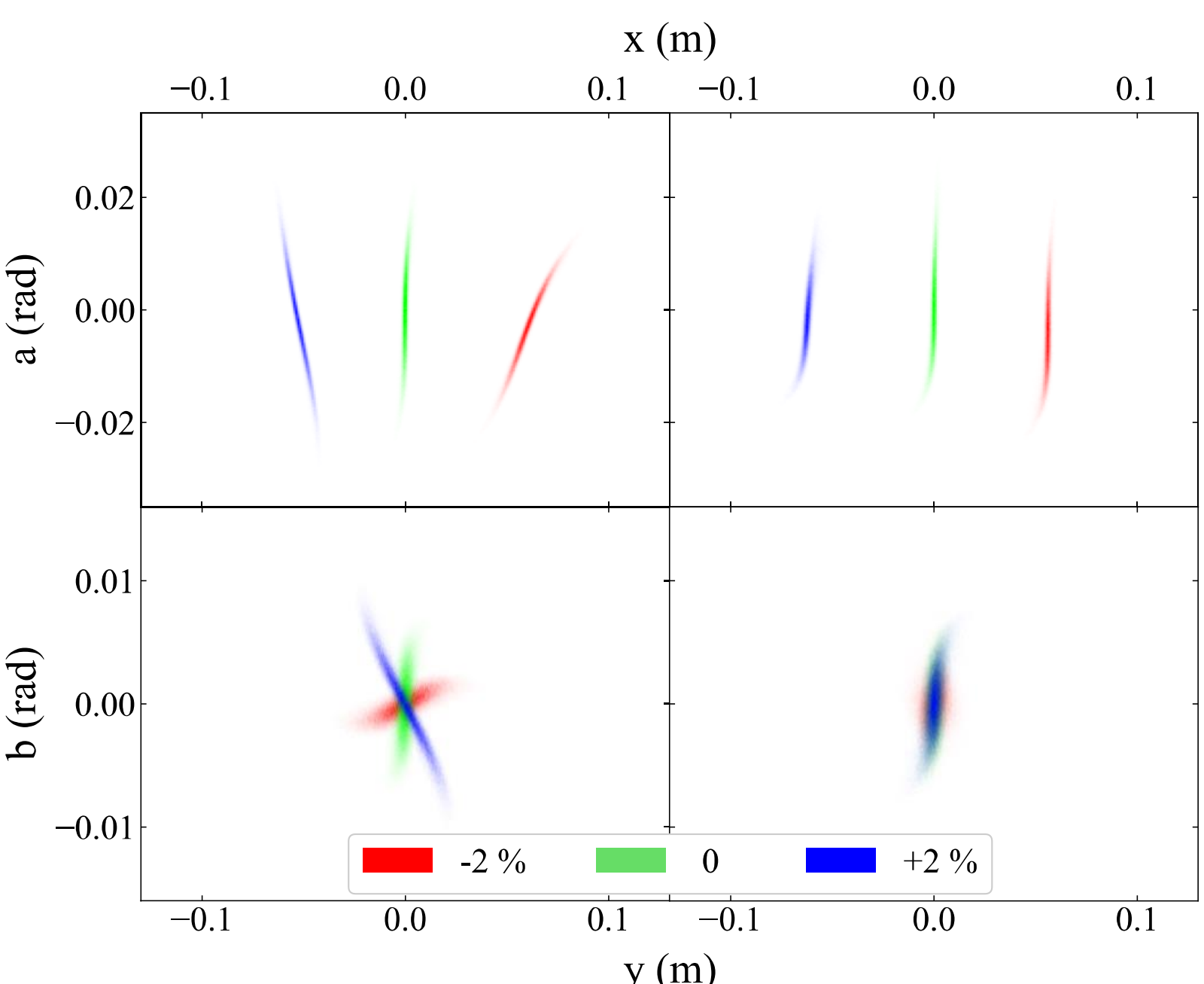

Figure 13. Beam phase space at PF2 focal plane without (left) and with (right) the third order aberration correction.

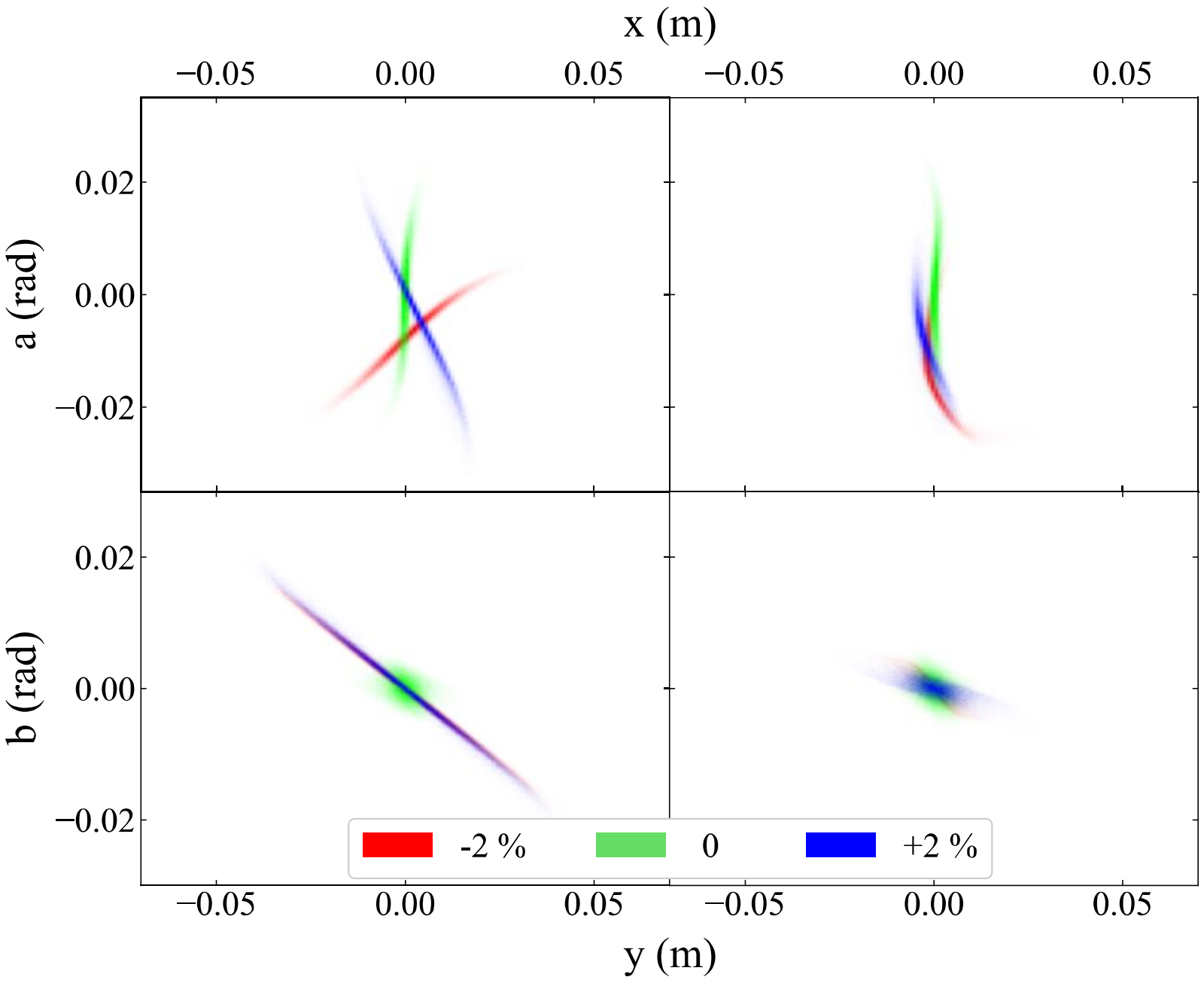

Figure 14. Beam phase space at PF4 focal plane without (left) and with (right) the third order

aberration correction.

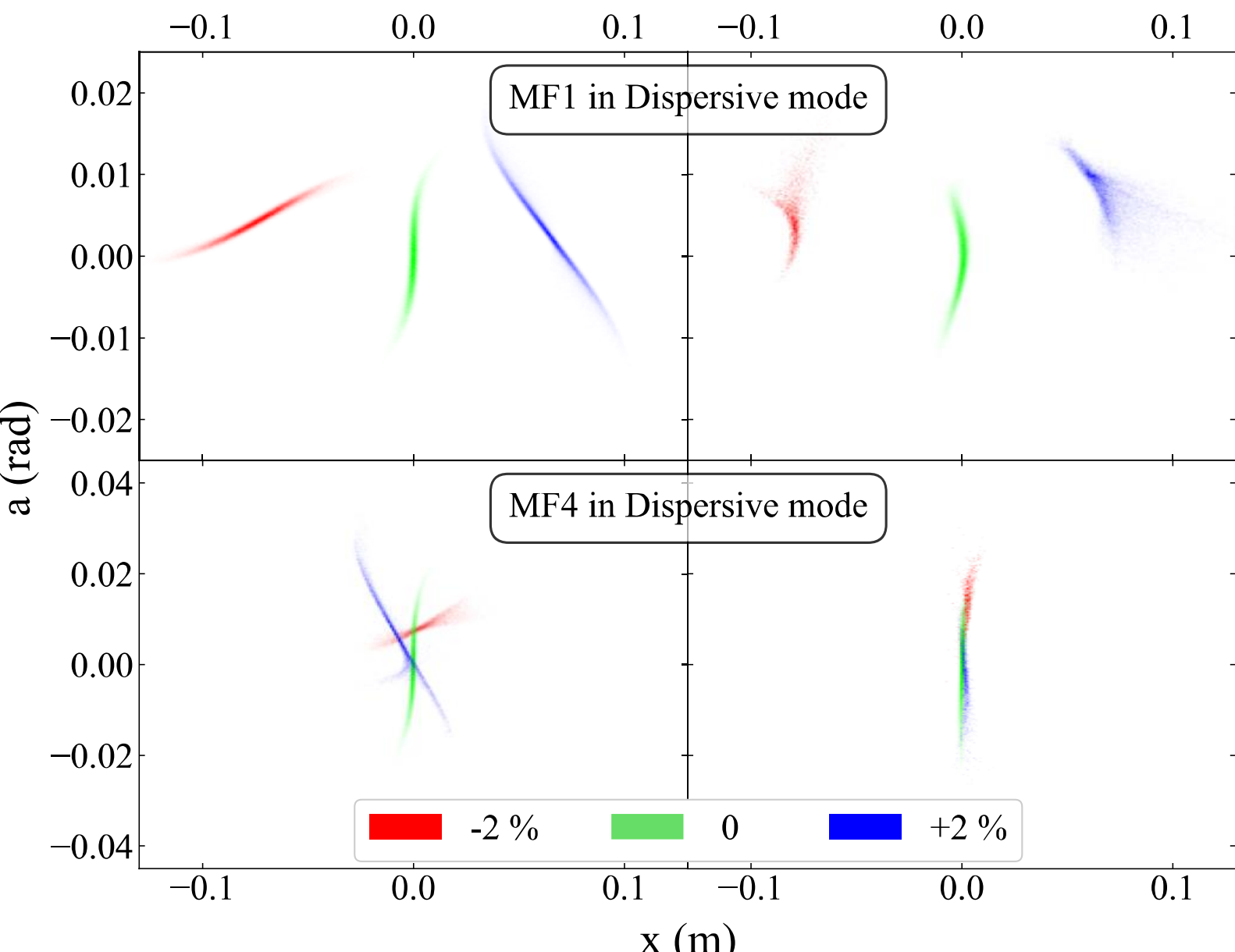

Figure 15. Beam phase space at MF1 and MF4 focal planes in dispersive mode without (left) and with (right) the third order aberration correction.

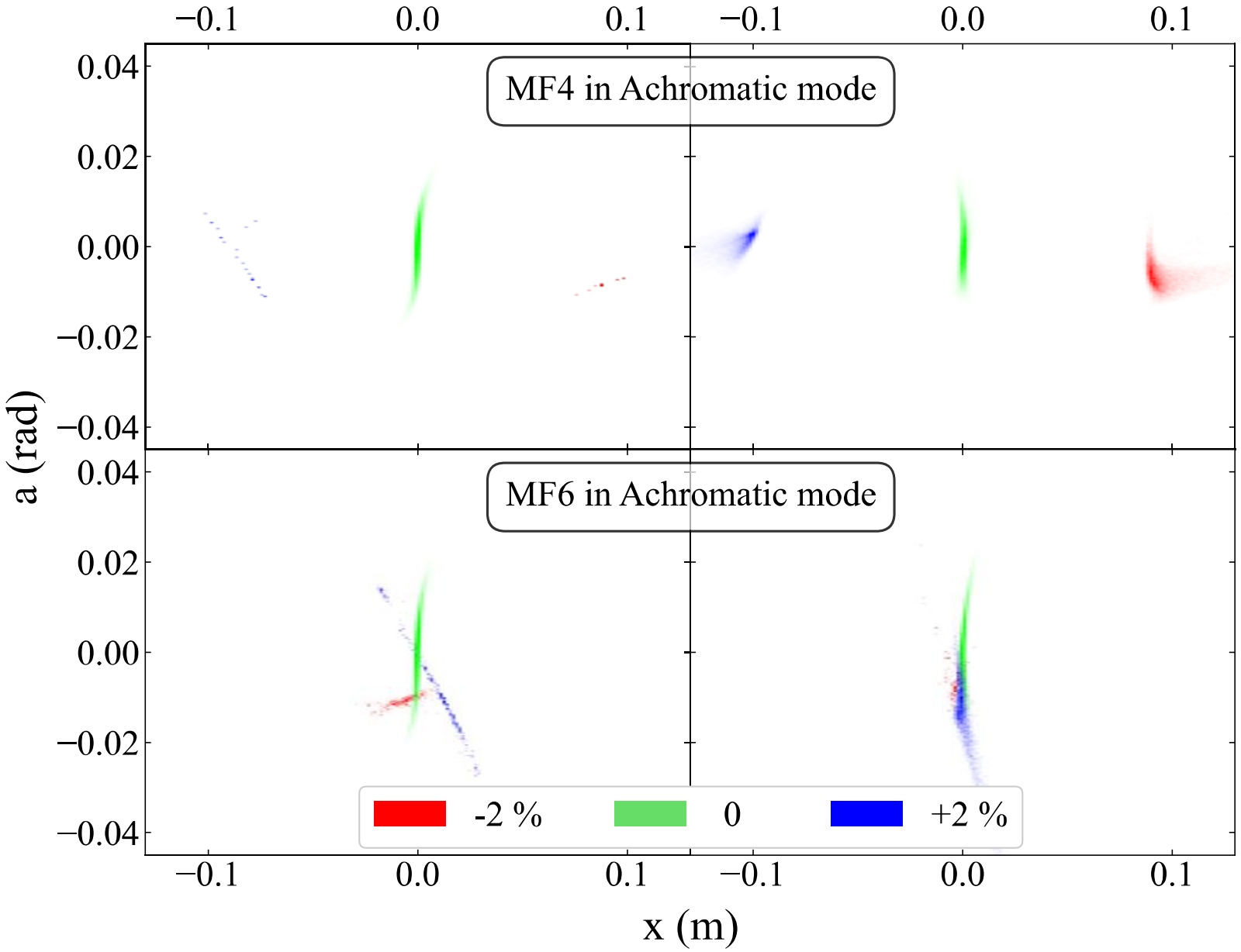

Figure 16. Beam phase space at MF4 and MF6 focal planes in achromatic mode without (left) and with (right) the third order aberration correction.

## 4. Performance evaluation

The evaluation of the experimental performance of HFRS Phase-I is carried out with the Monte Carlo simulation program MOCADI. The third order COSY maps are used in simulations. The apertures of multipole magnets, the gaps and good field regions of dipole magnets are all taken into account.

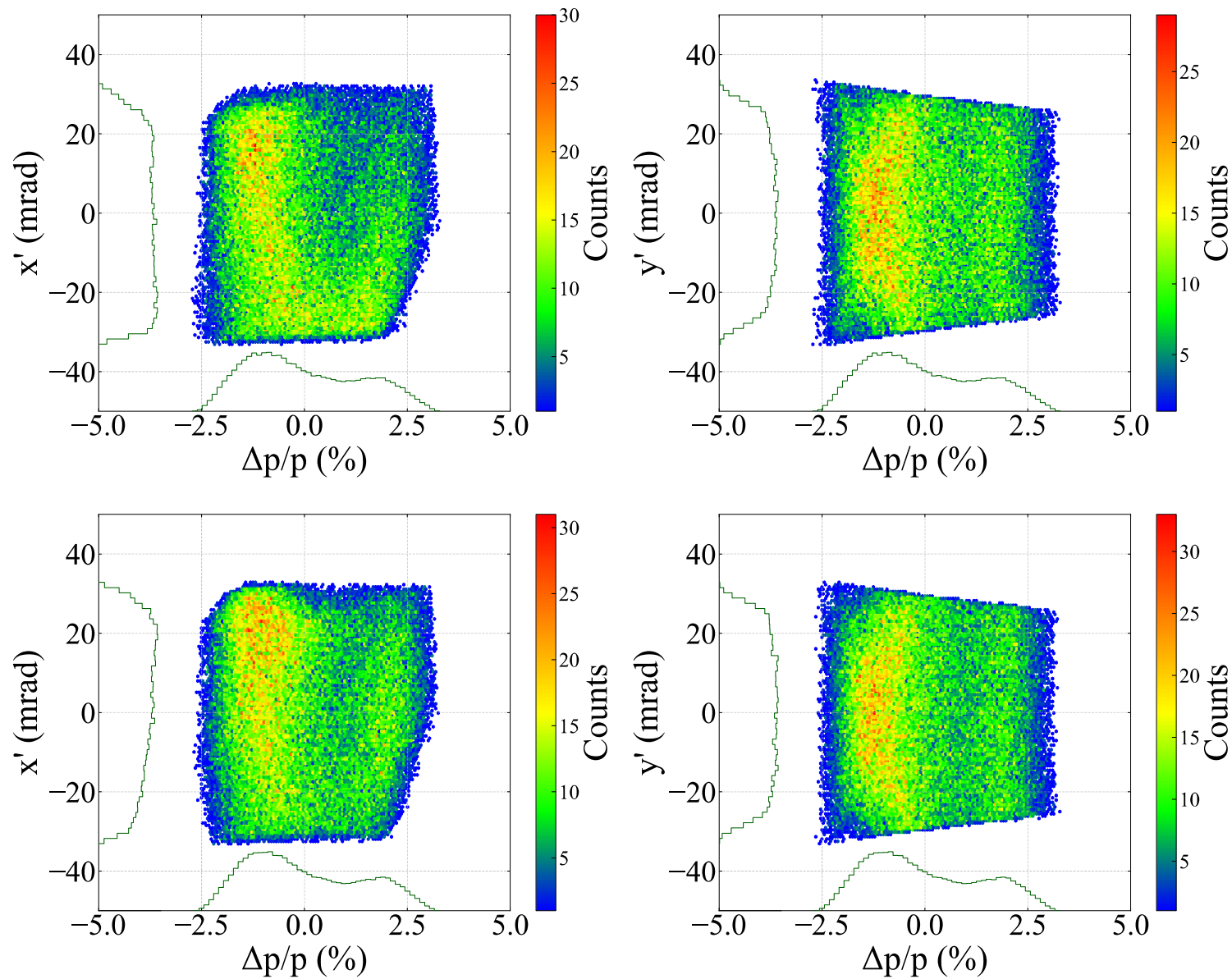

Figure 17. Momentum and angular acceptances for dispersive mode (upper) and achromatic mode (lower).

A beam with a large emittance and large momentum spread is injected into HFRS to evaluate its acceptance. When the beam is transported in HFRS, most of it is lost, and the surviving particles at the exit are those within the acceptance. Extracting the initial state of the surviving particles can yield the acceptance information. Figure 17 shows the momentum and angular acceptances in dispersive mode and achromatic mode. The acceptance is calculated to contain 95% of the survived particles. In dispersive mode, the momentum acceptance is about 2.4%, and the angular acceptances are about 29 mrad in horizontal direction and 27 mrad in vertical direction. In achromatic mode, the momentum acceptance is about 2.4%, and the angular acceptances are about 29 mrad in horizontal direction and 27 mrad in vertical direction. In both operation modes, both the momentum acceptance and the angular acceptances meet the design goals as described in Table 2.

As a doubly magic nucleus with exceptional doubly-closed-shell configuration [22], $^{132}$Sn has been a cornerstone in nuclear physics research. Its unique structural characteristics make it an ideal exotic nucleus for investigating nuclear interactions, structural evolution, and astrophysical nucleosynthesis processes. Fission populates a large phase space volume, which also makes it a good test case. Therefore, the $^{132}$Sn generated through $^{238}$U projectile fission will serve as a critical probe for evaluating the separation capability of the HFRS in achromatic mode. The doubly magic nucleus $^{100}$Sn, due to its extreme rarity, very-low production cross-section, and high difficulty in separation, is used to evaluate the separation capability of the HFRS in dispersive mode through the projectile fragmentation reaction of $^{124}Xe$.

Table 5. Parameters of Monte Carlo simulations.

| | | |
|---|---|---|
| Primary Ion | $^{238}U^{35+}$ | $^{124}Xe^{54+}$ |
| Energy [MeV/u] | 833 | 1098 |
| Emittance [π mm·mrad] | 1 (H) / 12 (V) | |
| Momentum Spread | 0.3 % | |
| Distribution Type | Gaussian | |
| Secondary Ion | $^{132}Sn^{50+}$ | $^{100}Sn^{50+}$ |
| Reaction Mechanism | Coulomb Fission<br>Abrasion-Fission | Projectile Fragmentation |
| Thickness of Graphite Target [mg/cm$^2$] | 1850 | 5000 |

The Monte Carlo simulation parameters are shown in the Table 5. The thickness of the graphite target is optimized. Three achromatic aluminum wedge degraders are employed to purify the exotic nucleus of interest. The first degrader is positioned at PF2. The second degrader is installed at MF1 and utilized in dispersive mode. The third degrader is placed at MF4 and employed in achromatic mode. The parameters of the three degraders are listed in Table 6.

Table 6. Parameters of degraders.

| Parameters | Degrader 1 | Degrader 2 | Degrader 3 |
|---|---|---|---|
| Thickness [mg/cm$^2$] | 6000 | 3500 | 3000 |
| Wedge Angle [mrad] | -14.19 ($^{100}Sn$)<br>-14.97 ($^{132}Sn$) | 7.58 ($^{100}Sn$) | -5.34 ($^{132}Sn$) |

Figure 18 shows the momentum and angular distributions of $^{132}Sn$ generated by the projectile fission of $^{238}U$. The angular distributions of $^{132}Sn$ ions are all within the angular acceptance range of HFRS. The momentum spread of a large number of particles exceeds 2.4%. It is predicted that quite much particles will be lost in the early transmission stage. Figure 19 shows the momentum and angular distributions of $^{100}Sn$ generated by the projectile fragmentation of $^{124}Xe$, which are mostly within the acceptance range of HFRS. The magnetic rigidity of the machine is set according to the central longitudinal momentum to accept as many secondary ions as possible. Figure 20 shows the separation of $^{132}Sn$ in achromatic mode (a) and $^{100}Sn$ in dispersive mode (b). As can be seen, both $^{132}Sn$ and $^{100}Sn$ are well separated and purified with two degraders separations.

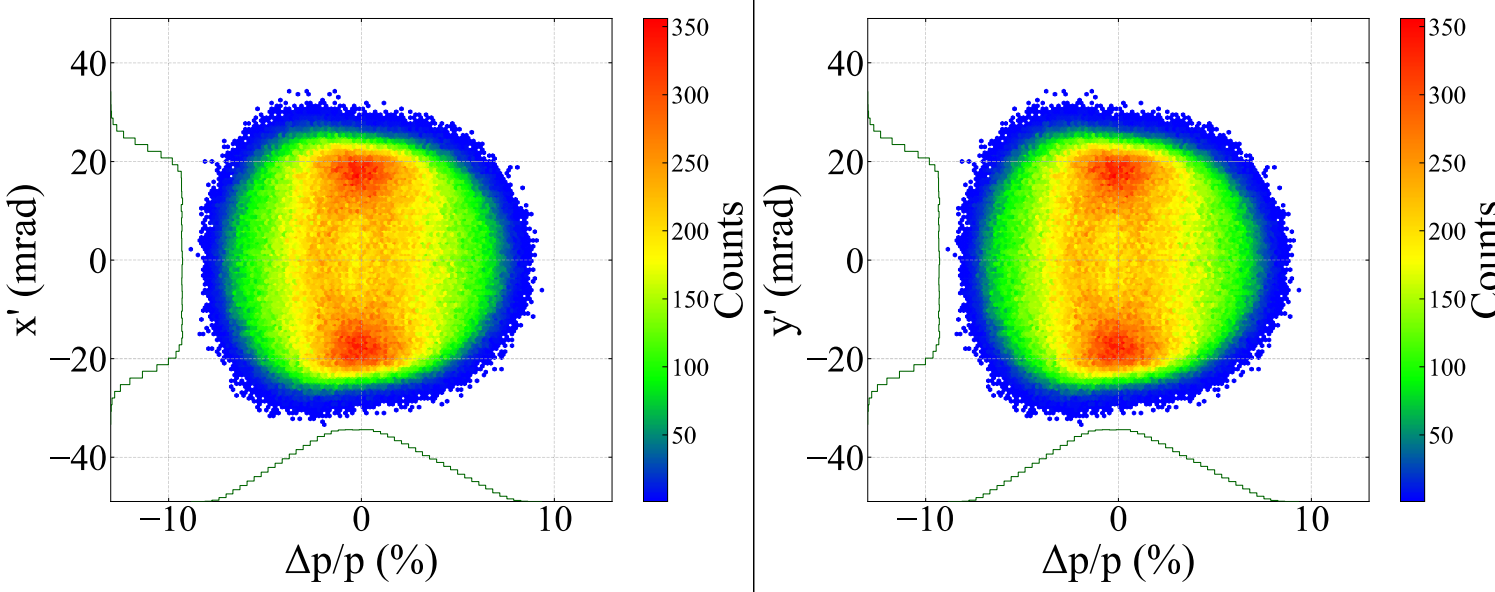

Figure 18. Momentum and angular distributions of $^{132}$Sn produced by projectile fission of $^{238}$U.

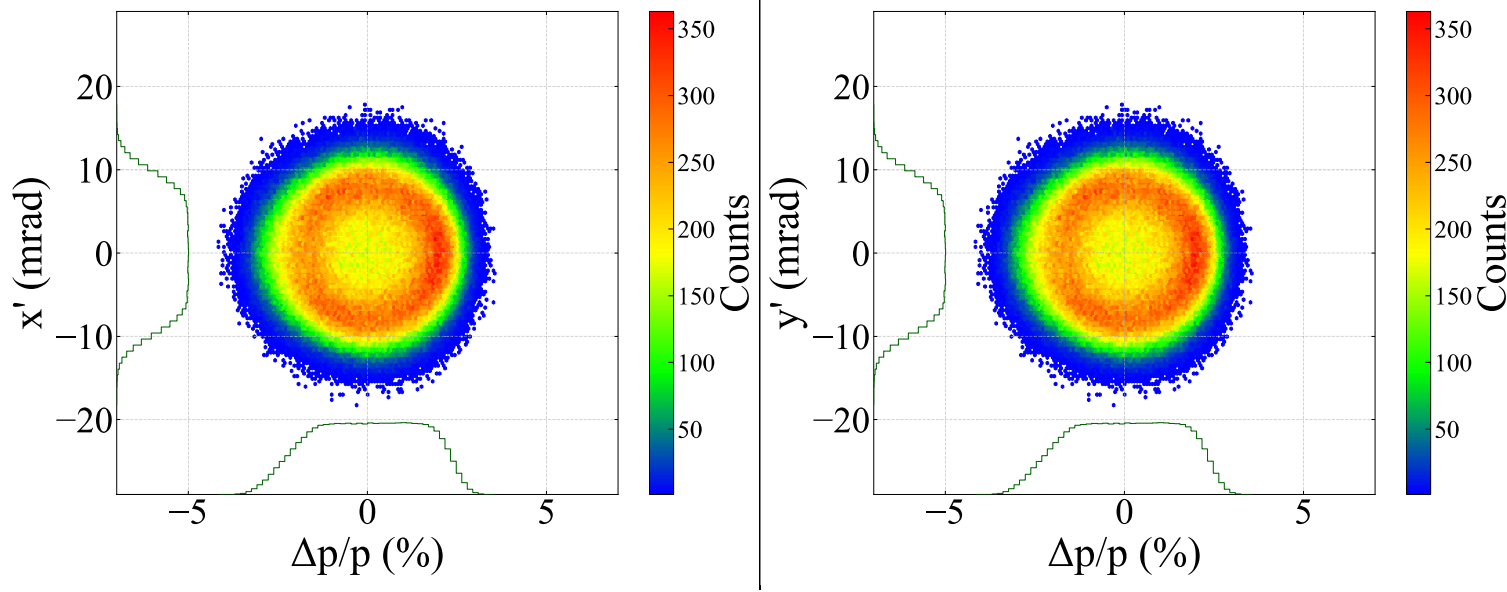

Figure 19. Momentum and angular distributions of $^{100}$Sn produced by projectile fragmentation of $^{124}$Xe.

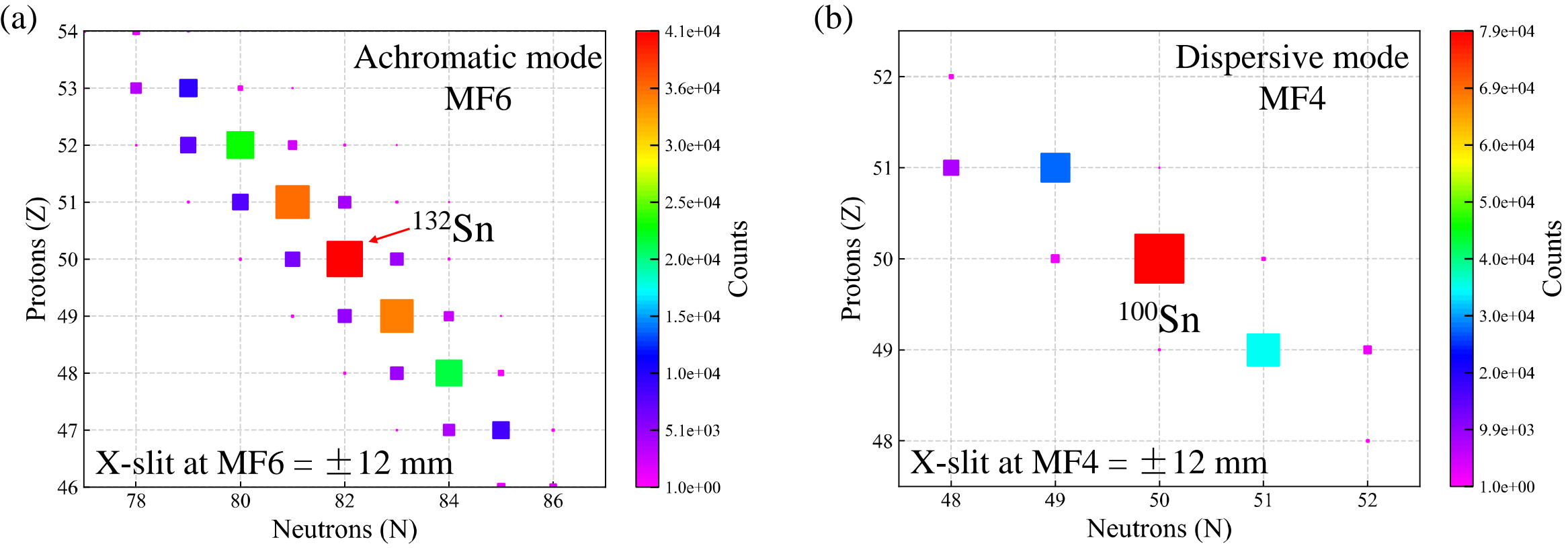

Figure 20. Separation of $^{132}$Sn in achromatic mode (a) and $^{100}$Sn in dispersive mode (b) with the third order optics.

The transmission efficiencies for $^{132}$Sn and $^{100}$Sn in the third order and the first order optics are shown in Fig. 21. Only ion-optical transmission losses are being calculated, and transmission losses due to the charge state change are not being considered. $^{132}$Sn experiences significant losses during the transfer from PF0 to PF2. This phenomenon is attributed to the large momentum spread of $^{132}$Sn. The large momentum spread of $^{132}$Sn is due to the kinematics of fission. The loss of $^{100}$Sn is relatively small, and it still maintains 80% transmission efficiency at the third order beam optics when reaching the final focal plane MF6.

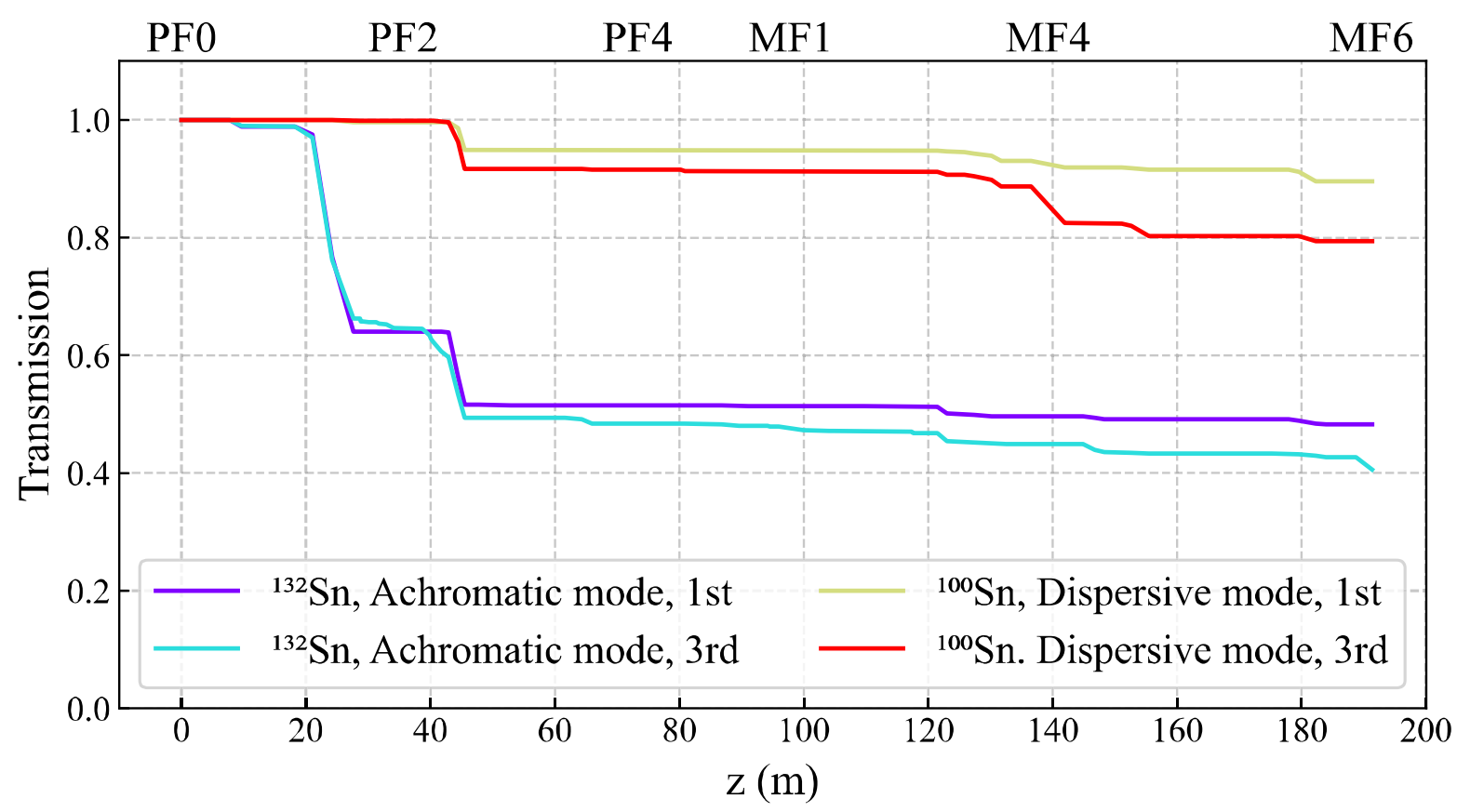

Figure 21. Transmissions of $^{132}$Sn and $^{100}$Sn in HFRS at the first and third order beam optics.

## 5. Summary

At present, all devices of HIAF have been installed. The joint debugging works of magnets and power supplies have also been completed. The beam commissioning will be carried out in this autumn.

In this paper, the measured data of the magnets are used to construct a high-precision first order and third order beam optics of HFRS Phase-I. A program based on COSY INFINITY/GICOSY, MOCADI, and Optuna has been developed, enabling very convenient correction of aberrations and higher order optics for HFRS. In both ion-optical modes, the momentum and angular acceptances meet the design goals. Finally, the performance of the separator is assessed by MOCADI code and two different reactions, projectile fragmentation reaction $^{124}$Xe+$^{12}$C→$^{100}$Sn and projectile fission reaction $^{238}$U+$^{12}$C→$^{132}$Sn. The simulation results demonstrate good separation capability and high transmission efficiency of the HFRS Phase-I, which can meet the current experimental requirements. The relevant results will be used for the subsequent beam commissioning of HFRS.

## Acknowledgements

This work is supported by High Intensity heavy-ion Accelerator Facility (HIAF) project approved by the National Development and Reform Commission of China.

## Conflict of interest statement

On behalf of all authors, the corresponding author states that there is no conflict of interest.

## References

[1] L.N. Sheng, X.H. Zhang, H. Ren, et al., Ion-optical updates and performance analysis of High energy FRagment Separator (HFRS) at HIAF, Nuclear Instruments & Methods in Physics Research Section B: Beam Interactions with Materials and Atoms, 547 (2024) 165214. doi: 10.1016/j.nimb.2023.165214.
[2] L.N. Sheng, X.H. Zhang, J.Q. Zhang, et al., Ion-optical design of High energy FRagment Separator (HFRS) at HIAF, Nuclear Instruments & Methods in Physics Research Section B: Beam Interactions with Materials and Atoms, 469 (2020) 1-9. doi: 10.1016/j.nimb.2020.02.026.
[3] X. Ma, W.Q. Wen, S.F. Zhang, et al., HIAF: New opportunities for atomic physics with highly charged heavy ions, Nuclear Instruments & Methods in Physics Research Section B: Beam Interactions with Materials and Atoms, 408 (2017) 169-173. doi: 10.1016/j.nimb.2017.03.129.
[4] J.C. Yang, J.W. Xia, G.Q. Xiao, et al., High Intensity heavy ion Accelerator Facility (HIAF) in China, Nuclear Instruments & Methods in Physics Research Section B: Beam Interactions with Materials and Atoms, 317 (2013) 263-265. doi: 10.1016/j.nimb.2013.08.046.
[5] X.H. Zhou, J.C. Yang, H.P. Team, Status of the high-intensity heavy-ion accelerator facility in China, AAPPS Bulletin, 32 (2022) 35. doi: 10.1007/s43673-022-00064-1.
[6] K.H. Schmidt, E. Hanelt, H. Geissel, et al., The momentum-loss achromat — A new method for the isotopical separation of relativistic heavy ions, Nuclear Instruments & Methods in Physics Research Section A: Accelerators Spectrometers Detectors and Associated Equipment, 260 (1987) 287-303. doi: 10.1016/0168-9002(87)90092-1.
[7] K. Wang, L.-N. Sheng, G. Wang, et al., High-precision beam optics calculation of the HIAF-BRing using measured fields, Journal of Instrumentation, 20 (2025) P08023. doi: 10.1088/1748-0221/20/08/P08023.
[8] S. Ruan, J. Yang, J. Zhang, et al., Design of extraction system in BRing at HIAF, Nuclear Instruments & Methods in Physics Research Section A: Accelerators Spectrometers Detectors and Associated Equipment, 892 (2018) 53-58. doi: 10.1016/j.nima.2018.02.052.
[9] W. Wu, E.M. Mei, W. You, et al., Superconducting Magnet Development for the HIAF Accelerator Complex, IEEE Transactions on Applied Superconductivity, 32 (2022) 4002305. doi: 10.1109/Tasc.2022.3153428.
[10] H. Grote, F. Schmidt, Mad-X - An upgrade from MAD8, Proceedings of the 2003 Particle Accelerator Conference, Vols 1-5, (2003) 3497-3499. doi: 10.1109/PAC.2003.1289960.
[11] M. Berz, Computational Aspects of Optics Design and Simulation - Cosy Infinity, Nuclear Instruments & Methods in Physics Research Section A: Accelerators Spectrometers Detectors and Associated Equipment, 298 (1990) 473-479. doi: 10.1016/0168-9002(90)90649-Q.
[12] M. Berz, The Code Cosy Infinity, Conference Record of the 1991 Ieee Particle Accelerator Conference, Vols 1-5, (1991) 354-356. doi: Doi 10.1109/Pac.1991.164299.
[13] H. Wollnik, B. Hartmann, M. Berz, Principles of GIOS and COSY, AIP Conference Proceedings, 177 (1988). doi: 10.1063/1.37817.
[14] N. Iwasa, H. Geissel, G. Münzenberg, et al., MOCADI, a universal Monte Carlo code for the transport of heavy ions through matter within ion-optical systems, Nuclear Instruments & Methods in Physics Research Section B: Beam Interactions with Materials and Atoms, 126 (1997) 284-289. doi: 10.1016/S0168-583X(97)01097-5.

[15] N. Iwasa, H. Weick, H. Geissel, New features of the Monte-Carlo code MOCADI, Nuclear Instruments & Methods in Physics Research Section B: Beam Interactions with Materials and Atoms, 269 (2011) 752-758. doi: 10.1016/j.nimb.2011.02.007.

[16] M. Berz, K. Makino, COSY INFINITY 10.2 Beam Physics Manual, 2023.

[17] H.A. Enge, Effect of Extended Fringing Fields on Ion-Focusing Properties of Deflecting Magnets, Review of Scientific Instruments, 35 (1964) 278-287. doi: 10.1063/1.1718806.

[18] T. Stuart, CAS - CERN Accelerator School : Magnetic measurement and alignment: Montreux, Switzerland 16 - 20 Mar 1992, CERN1992.

[19] M. Berz, H.C. Hoffmann, H. Wollnik, COSY 5.0 — The fifth order code for corpuscular optical systems, Nuclear Instruments & Methods in Physics Research Section A: Accelerators Spectrometers Detectors and Associated Equipment, 258 (1987) 402-406. doi: 10.1016/0168-9002(87)90920-X.

[20] T. Akiba, S. Sano, T. Yanase, et al., Optuna: A Next-generation Hyperparameter Optimization Framework, Proceedings of the 25th ACM SIGKDD International Conference on Knowledge Discovery & Data Mining, Association for Computing Machinery, Anchorage, AK, USA, 2019, pp. 2623–2631. doi: 10.1145/3292500.3330701.

[21] SQLite. https://www.sqlite.org/

[22] T. Björnstad, M.J.G. Borge, J. Blomqvist, et al., The doubly closed shell nucleus 13250Sn82, Nuclear Physics A, 453 (1986) 463-485. doi: 10.1016/0375-9474(86)90447-1.